\documentclass[a4paper,10pt,ps2pdf]{article}

\usepackage[numbers]{natbib}
\usepackage{tocbibind} 
\settocbibname{References}

\usepackage{amssymb} 
\usepackage{amsmath} 

\usepackage{cancel}

\usepackage{hyperref}
\newcommand{\td}{\mathrm{d}}
\newcommand{\te}{\mathrm{e}}
\newcommand{\ti}{\mathrm{i}}

\usepackage{graphicx}
\usepackage{subfigure}
\usepackage{psfrag}
\usepackage{color}

\graphicspath{{./img/}{./}}

\begin{document}
\title{Electrodynamics of Abrikosov vortices: the Field Theoretical Formulation}
\author{A.J. Beekman$^{1}$ and J. Zaanen$^{1}$}


\maketitle
\noindent ${^1}$Instituut-Lorentz for Theoretical Physics, Universiteit Leiden, P.O. Box 9506, 2300 RA Leiden, The Netherlands

\begin{abstract}
 Electrodynamic phenomena related to vortices in superconductors have been studied since their prediction by Abrikosov, and seem to hold no fundamental mysteries. However, most of the effects are treated separately, with no guiding principle. We demonstrate that the relativistic vortex worldsheet in spacetime is the object that naturally conveys all electric and magnetic information, for which we obtain simple and concise equations. Breaking Lorentz invariance leads to down-to-earth Abrikosov vortices, and special limits of these equations include for instance dynamic Meissner screening and the AC Josephson relation. On a deeper level, we explore the electrodynamics of two-form sources in the absence of electric monopoles, in which the electromagnetic field strength itself acquires the characteristics of a gauge field. This novel framework leaves room for unexpected surprises.
\end{abstract}

\section{Introduction.}

The study of the matter formed from Abrikosov vortices in type-II su\-per\-con\-duc\-tors\cite{Abrikosov57}  constitutes a vast and mature research subject. This subject is crucial for the technological  applications of superconductivity\cite{RogallaKes11} but it has also proven to be a fertile source for fundamental condensed matter physics research. The elastic and hydrodynamical properties of matter formed from vortices can be  very easily tuned by external means and it has demonstrated to be an exceedingly fertile model system to study generic questions regarding crystallization, the effects of background quenched disorder and so forth\cite{Nelson88, RosensteinLi10}. Especially after the discovery of the cuprate high-$T_\text{c}$ superconductors it became also possible to study the fluids formed from vortices. Because of the strongly two-dimensional nature of the superconductivity in the cuprates, the Abrikosov vortex lattice becomes particularly soft and it melts easily due to thermal motions at temperatures that are  much below the mean field  $H_{c2}$-line \cite{BlatterEtAl94}. 

Many phenomena in this field are of a dynamical nature, associated with the fact that vortices are in motion. This includes the vortex flow, the magnetic field penetration and the flux creep, but also the large Nernst effect of the vortex fluid and, perhaps most spectacularly, the use of cuprate vortices as source of terahertz radiation\cite{BulaevskiiChudnovsky06,BulaevskiiKoshelev06}. This vortex dynamics is analogous to the magnetohydrodynamics of electrically charged plasmas in the sense that the forces exerted on vortices are exclusively of  electromagnetic origin, while in turn the vortex matter backreacts on the electromagnetic fields. The phenomena that arise are rather thoroughly understood departing from the AC-and DC-Josephson relations as well as the  Maxwell equations as the force equations in this ``vortex magneto-hydrodynamics''. 

Although the computations explaining these phenomena are correct, they are of a rather improvised ad hoc nature, at least compared to the Landau--Lifshitz style\cite{LandauLifshitz60} of deriving the usual magnetohydrodynamics from first principles. In so far as the forces are concerned, the latter eventually departs from the microscopic action describing  Maxwell electrodynamics,
 \begin{align}
 S & = \int \td^4 x\  \mathcal{L}_\text{Maxwell}, \\
 \mathcal{L}_\text{Maxwell}  & =   -\frac{1}{4} F_{\mu\nu} F_{\mu\nu} + A_\mu J_\mu,
 \label{eq:Maxwell action}
\end{align}
where the conserved currents $J_{\mu}$ parametrize the world lines of the charged particle forming the plasma, sourcing via the electromagnetic gauge fields $A_{\mu}$ the electromagnetic field strength $F_{\mu\nu} = \partial_\mu A_\nu - \partial_\nu A_\mu$. Varying the action with respect to $A_\nu$ one obtains the Euler--Lagrange equations of motion,
\begin{equation}
 \partial_\mu F_{\mu\nu} = -J_\nu,\label{eq:Maxwell equations} 
\end{equation}
the Maxwell equations in relativistic shorthand notation. What are the corresponding equations, dealing with the Abrikosov vortices sourcing the electromagnetic fields? In a related but yet different context\cite{BeekmanZaanen11b} we stumbled on this question, finding out that the answer is apparently not available in the literature. To our impression this is rooted in the fact that one needs a piece of mathematical technology that is unfamiliar to  condensed matter physicists, while it is well known in differential geometry\cite{HenneauxTeitelboim92} and string theory\cite{Polchinski98}.  We refer to the formalism of two-form gauge theory. This amounts to a generalization of the familiar ``one-form'' Maxwell (and Yang--Mills) gauge theories characterized by gauge fields (vector potentials) $A_\mu$ that carry one label. Instead one has gauge fields $b_{\mu \nu}$, representing antisymmetric tensors, that have gauge transformations in the form of the addition of the gradient of a vector field (that therefore looks like a Maxwell field strength $f_{\mu\nu} = \partial_\mu a_\nu - \partial_\nu a_\mu$ even though the meaning is completely different): $b_{\mu \nu} \rightarrow b_{\mu \nu} + f_{\mu \nu}$. How can such a structure
arise? 

The key ingredient is that vortices as topological excitations of the superconductor are lines in 3-dimensional space when they are static. Invoking time as one should using the action principle, these lines spread out in space time as surfaces or {\em world sheets}, in the same way as point particles correspond with world lines. In fact, in string theory, Abrikosov vortices are known as ``Nielsen--Olesen strings''\cite{NielsenOlesen73}, regarded as rather primitive as compared to the fundamental ``critical'' strings. We can now rely on Schwinger's principle \cite{Schwinger70} that the sources are the principal objects, emitting and absorbing the force-mediating gauge fields. Dealing with point particles as the sources, their infinitesimal world line intervals are parametrized in terms of a charge density $\rho$ and current $\mathbf{J}$ which can be combined in the spacetime covariant current $J_\mu = ( c\rho , \mathbf{J})$ of  Eq.'s \eqref{eq:Maxwell action} and \eqref{eq:Maxwell equations}, that satisfies a continuity equation,
\begin{equation}
 \partial^\mu J_\mu = \partial_t \rho  + \nabla \cdot \mathbf{J} = 0.
\end{equation}
The continuity equation is also a constraint that removes the longitudinal component of $J_\mu$. The fields $A_\mu$ that couple to $J_\mu$ are therefore gauge fields, since the longitudinal component is not sourced and as a consequence not physical. In this context of electromagnetic currents, the field $A_\mu$ just corresponds with the photon gauge field. 

In the case of the world sheets formed by the vortices in 3+1-dimensional spacetime one has to depart instead from infinitesimal world sheet areas and these have to be parametrized in antisymmetric tensors $J^\text{V}_{\mu\nu}$. In analogy with the particle currents, these ``two-form currents'' will source two-form gauge fields, subjected to the transversality condition. The vortex action has therefore to contain a term $ b_{\mu \nu} J_{\mu \nu}$ where $b_{\mu \nu}$ is a two-form gauge field. As we will show, for the vortices in 3+1 dimensions the pieces fit in a remarkably elegant way:  instead of the gauge potential $A_{\mu}$,  now the physical electric and magnetic fields $\mathbf{E}$ and $\mathbf{B}$ as collected in the Maxwell field strength $F_{\mu\nu}$, couple minimally as two-form gauge fields  to the vortex world sheet current. As we will show, this gauge coupling involves  actually  the  (Hodge) dual of the electromagnetic field strength, so that the coupling term is  $F_{\mu\nu} \epsilon^{\mu\nu\kappa\lambda} J^\text{V}_{\kappa\lambda}$, where $\epsilon^{\mu\nu\kappa\lambda}$ is the fully antisymmetric Levi-Civita symbol. Remarkably, in so far this ``Schwinger side'' is concerned, the one-form Maxwell theory is lifted by an emergent gauge invariance to a two-form gauge theory. Given that inside the type-II superconductor only vortices are present as sources for the electromagnetic fields, and since these carry only magnetic fluxes and no electrical monopole charges, the Maxwell field strengths of the fundamental vacuum lose their physical meaning and turn into gauge fields.  However, there is a caveat: besides the vortices the electromagnetic fields also experience the Meissner- and Thomas--Fermi screening. These restore the physical reality of the Maxwell field strengths, explicitly  breaking the two-form gauge invariance. Only in the extreme type-II limit, where both the Meissner and Thomas-Fermi screening lengths diverge, the emergent two-form gauge invariance becomes literal. Otherwise the two-form gauge theory is forced in a ``fixed frame'' corresponding with the generalized Lorenz gauge fix of the fully gauge invariant theory. This peculiar gauge theoretical structure is the key to the special nature of vortex electrodynamics. 

On the practical side, with this mathematical technology at hand it is straightforward to derive the fundamental action and equations of motion governing the electrodynamics of vortices in type-II superconductors: Eqs. \eqref{eq:relativistic SC EoM} and \eqref{eq:magnetic equation}, \eqref{eq:longitudinal electric equation} \& \eqref{eq:transversal electric equation} for the simpler relativistic- and the realistic non-relativistic  vortices, respectively. These originate from an action Eq. \eqref{eq:two-form action with Meissner} resp. \eqref{eq:non-relativistic vortex action}. Merely for reasons of convenience we will develop much of the theory in the relativistic limit where it is assumed that the superfluid phase velocity is coincident with the speed of light. As in electrodynamics, this greatly simplifies the equations and it is easier to follow the argument. Of course, in real superconductors this phase velocity is a small fraction of the light velocity  and it is straightforward to break the Lorentz invariance to obtain the equations presented in Section  \ref{sec:vortex electrodynamics} which are of relevance to earthly superconductors.  

We claim that these equations of motion represent a  fundamental and  complete description of the  electrodynamics of the  Abrikosov vortices in superconductors. We will demonstrate that all known physical properties associated with this electrodynamics are economically encoded in  this formalism:  Thomas--Fermi screening and dynamical Meissner screening, the  electric fields induced by vortex motion and the Nernst effect, the AC Josepson effect and radiation due to vortex motion. These all arise as special limits of our general equations. Admittedly we have not achieved  extracting new vortex physics but we believe that there is a potential for surprises. We invite the expert readership to have a closer look to find out whether they can profit from this particularly convenient  mathematical formulation of the problem.  
 
The paper is organized as follows: the fundamental definition of a vortex is reviewed in Section \ref{sec:vortex fields}, and it is shown how vortices are described as world lines and world sheets. In Section \ref{sec:vortex world sheet in superconductors} we show how, in a superconductor, the vortex world sheet dynamics  follows directly from the time-dependent Ginzburg--Landau equations by a single mathematical derivative operation. This is the easiest way to actually obtain our central results. The next two sections have a theoretical emphasis: in Section \ref{sec:electrodynamics of two-form sources} we motivate the general mathematical structure of the electrodynamics sourced by two-form fields. We demonstrate  here  that the field strength $F_{\mu\nu}$ becomes itself a gauge field. A rigorous derivation of the action governing the electrodynamics of vortex sources using duality techniques is presented in Section \ref{sec:vortex duality in charged superfluids}. This comprises the dual gauge field that mediates interactions between vortices. Finally, by breaking the Lorentz invariance  we derive the equations of motion of vortex electrodynamics from this action in their full extent in Section \ref{sec:vortex electrodynamics}, and we derive the physical phenomena mentioned in the previous paragraph. Finally, in the Appendix  this structure is formulated in the  general  mathematical language  of differential forms,  allowing  us to extend the results to any spatial dimension larger than 2.

\section{Vortex fields.}\label{sec:vortex fields}

Let us  consider vortices in type-II superconductors. These are of course the familiar flux lines where the magnetic field penetrates through the superconducting sample, but the more profound statement is that a vortex is a region where the superconducting phase is singular. Let us repeat precisely what is meant by this  for the case of a 2+1-dimensional superconductor. This argument applies just as well to a superfluid  (uncharged superconductor). More details are found in Kleinert's textbook \cite[ch. II.1]{Kleinert89a}.

\begin{figure}
\hfill
 \subfigure[Phase coherence. {\scriptsize Spontaneous symmetry breaking singles out a preferred phase value.}]{\includegraphics[height=4cm]{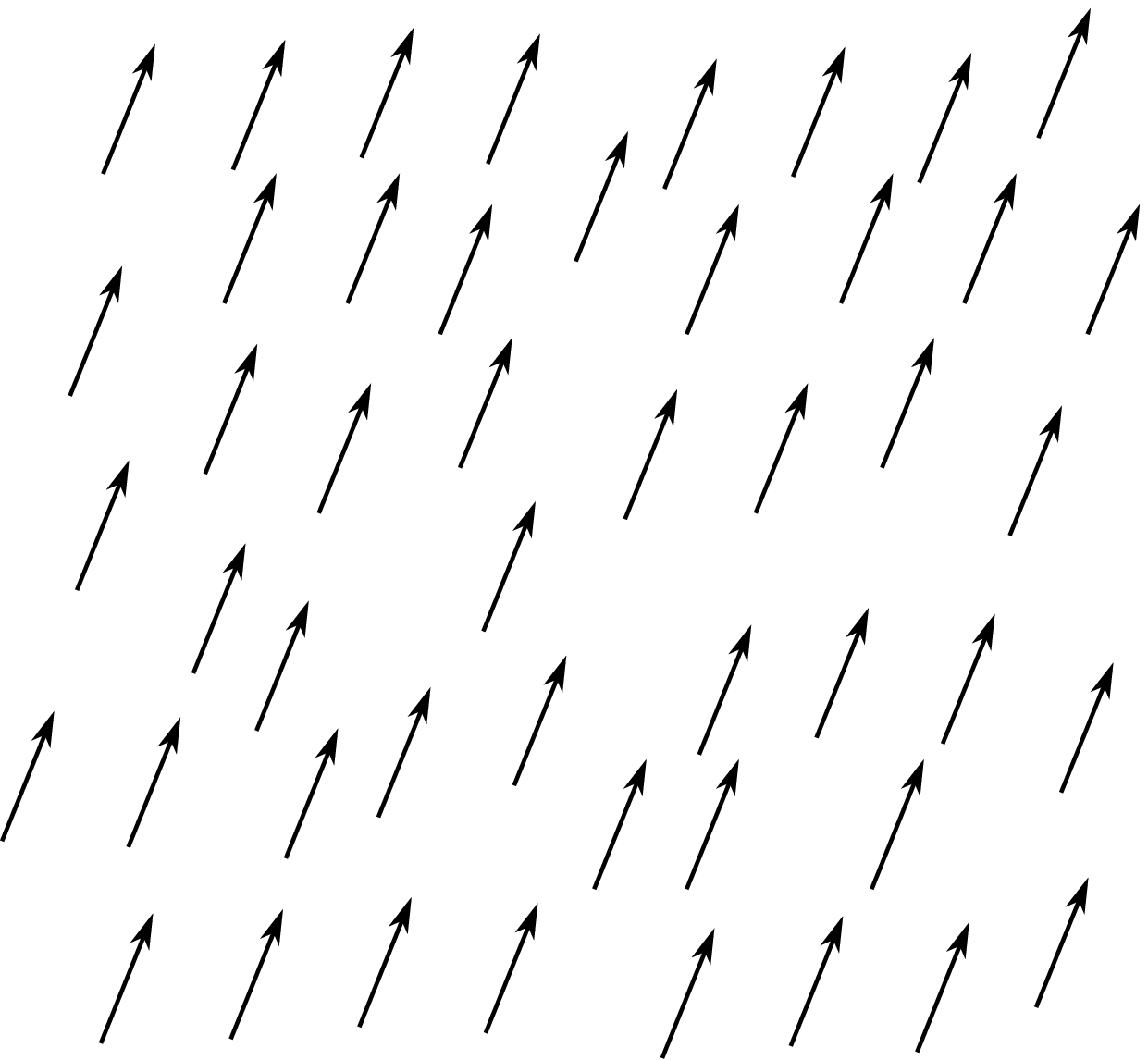}\label{fig:phase coherence}}
 \hfill
 \psfrag{c}{{\color{red}$\mathcal{C}$}}
 \subfigure[Vortex. {\scriptsize A single vortex of winding number 1. Tracking the phase change along the contour $\mathcal{C}$ (red) adds up to $2\pi$. Also shown in hatched red is the surface $\mathcal{S}$ enclosed by $\mathcal{C}$.}]{\includegraphics[height=4cm]{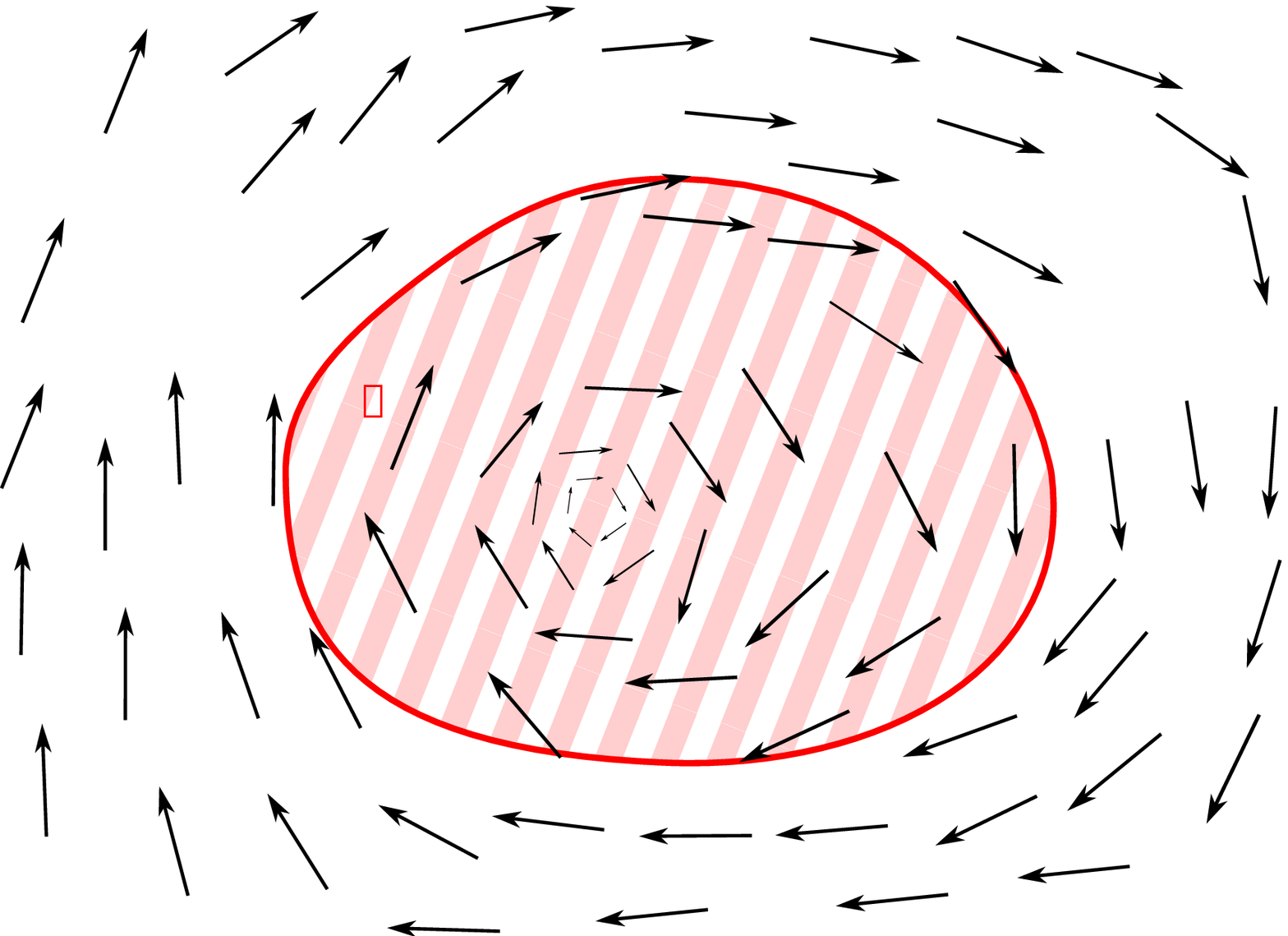}\label{fig:single vortex}}
 \hfill
\caption{Phase coherence and its distortion by a vortex. {\scriptsize The hallmark of the superconductor  is the long-range correlations in the order parameter phase. The phase is a compact variable, and as such permits vortex configurations, that cause the phase to change along its perimeter in multiples of $2\pi$ only. }}
\end{figure}

In the Ginzburg--Landau description, the parent state of a superconductor obeys a $U(1)$ symmetry, as the phase $\varphi$ of the complex order parameter $\Psi = \lvert \Psi \rvert \te^{\ti \varphi}$ is only defined up to global\footnote{In a superconductor this is lifted to the local $U(1)$ gauge symmetry of electromagnetism, but that does not matter  here as is well known. As we will see explicitly, the gauge symmetry can be included in a next step.} 
transformations $\Psi \to \Psi \te^{\ti \alpha}$. However, across the phase transition, the amplitude obtains a vacuum expectation value, and one of all possible phases is singled out. This is the well known phenomenon of spontaneous symmetry breaking. Hence, the superconductor is a state where the phase has a preferred value over large length scales (Fig. \ref{fig:phase coherence}). It is now possible that the system is disturbed so that a vortex is formed. In two spatial dimensions, a vortex is pointlike (a `vortex pancake'). Tracking the deviation of the phase from its preferred value along a contour $\mathcal{C}$ around the vortex origin reveals that the phase changes by $2\pi N$, where $N$ is the winding number (Fig. \ref{fig:single vortex}). The mathematical statement is,
\begin{equation}
 \oint_\mathcal{C} \td x_\mu\ \partial_\mu \varphi = 2\pi N.
\end{equation}
We can now formally invoke Stokes' theorem, by converting the contour integral into a surface integral of the curl of the integrand, over the surface $\mathcal{S}$ enclosed by $\mathcal{C}$,
\begin{equation}
 \int_\mathcal{S} \td S_\lambda \epsilon_{\lambda \nu \mu}\ \partial_\nu \partial_\mu \varphi =  2\pi N.
\end{equation}
This is allowed rigorously only when $\varphi$ is non-singular in all of $\mathcal{S}$, but we turn necessity into virtue by noticing that the above equation is satisfied when,
\begin{equation}\label{eq:vortex delta function}
 \epsilon_{\lambda \nu \mu} \partial_\nu \partial_\mu \varphi(x) = 2\pi N \delta^{(2)}_\lambda (x).
\end{equation}
Here $\delta^{(2)}_\lambda (x)$ is a two-dimensional delta-function in the plane orthogonal to $\lambda$. It is only non-zero at the origin. This corresponds to the fact that for a vortex $\varphi$ is singular at the origin. For the same reason, the derivatives on the left-hand side do not commute and the expression is not manifestly zero, as one would expect for the contraction with the fully antisymmetric tensor $ \epsilon_{\lambda \nu \mu}$.

\begin{figure}\label{figure2}
\hfill
 \psfrag{t}{$t$}
\psfrag{x}{$x$}
\psfrag{y}{$y$}
\psfrag{m}{$J_\mu$}
\psfrag{i}{$J_t$}
\psfrag{n}{$\mathbf{J}$}
 \subfigure[Vortex world line. {\scriptsize Segment of a vortex world line (red) in 2+1 dimensions. Shown in blue is the line element $J_\mu$ and it temporal and spatial components $J_t$ and $\mathbf{J}$. For an electric monopole, these correspond to the charge and current densities $\rho$ and $\mathbf{J}$.}]{\includegraphics[height=3.5cm]{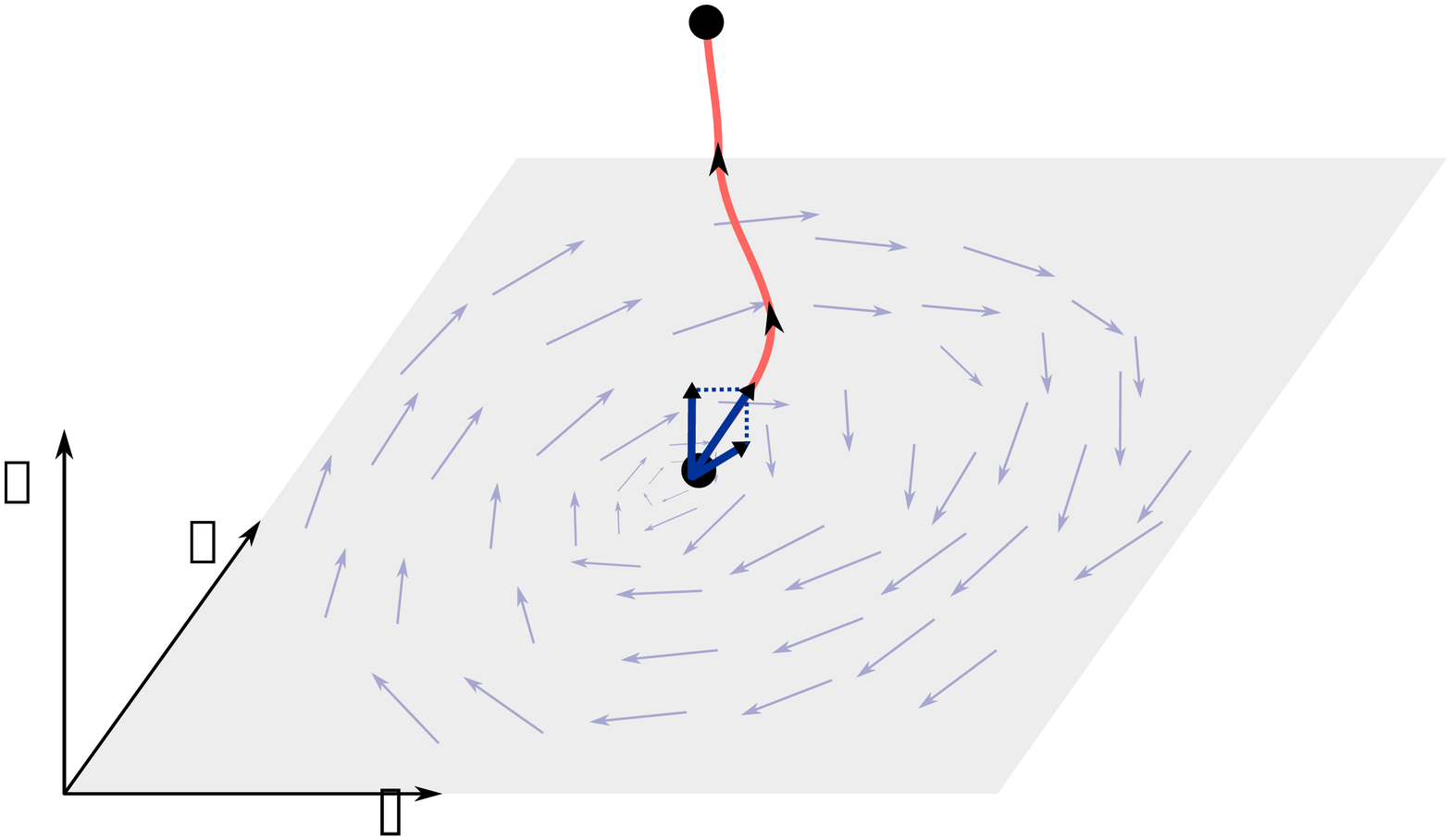}\label{fig:single vortex world line}}
 \hfill
\psfrag{t}{$t$}
\psfrag{x}{$x$}
\psfrag{y}{$y$}
\psfrag{k}{$\kappa$}
\psfrag{l}{$\lambda$}
 \subfigure[Vortex world sheet. {\scriptsize Segment of a vortex line tracing out a world sheet in time. The vortex lies in the $x$-$y$-plane, the third spatial dimension cannot be drawn. Shown in blue is the world sheet surface element $J_{\kappa\lambda}$.}]{\includegraphics[height=3.5cm]{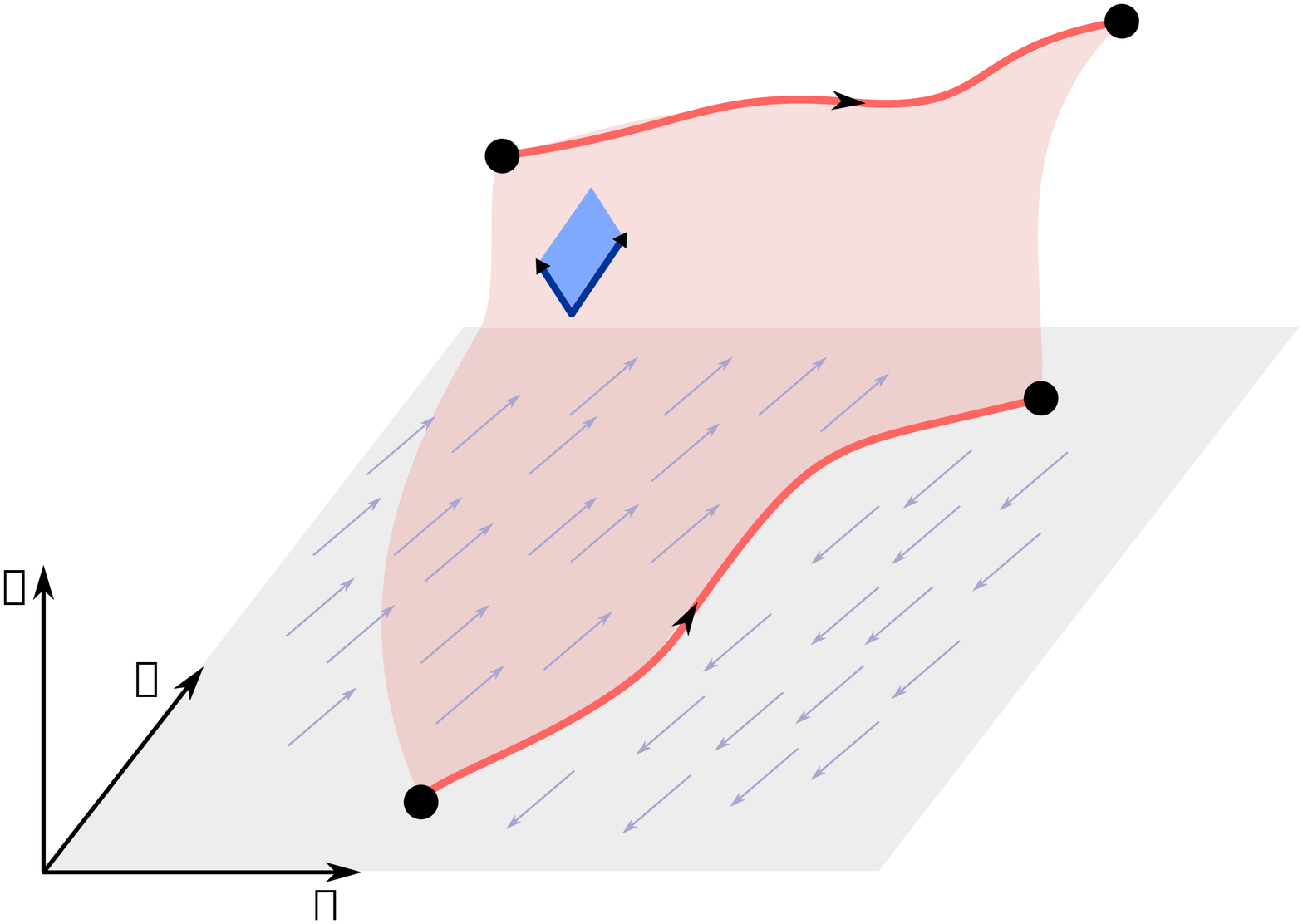}\label{fig:single vortex world sheet}}
 \hfill
\caption{Vortex world line and world sheet {\scriptsize A graphical picture of dynamical system is provided by tracing the motion of particles or line objects through time. The traces are called world lines or world sheets. In field theory, they are represented by one- or two-form fields, which represent the line or surface elements of the world line or sheet. If the world lines or sheets can not end in the middle of a system, the one- or two-form fields are conserved, as $\partial_\mu J{\mu} = 0$  or $\partial_\kappa J_{\kappa\lambda} = 0$.}}
\end{figure}

We define,
\begin{equation}\label{eq:2+1D vortex current definition}
J^\text{V}_\lambda(x) =  \epsilon_{\lambda \nu \mu} \partial_\nu \partial_\mu \varphi(x), 
\end{equation}
as the \emph{vortex current}. Its temporal component,
\begin{equation}\label{eq:vortex current delta function}
 J^\text{V}_t (x) = (\partial_x \partial_y - \partial_y \partial_x) \varphi(x) = \delta^{(2)} (\mathbf{x}),
\end{equation}
is the density of vorticity, just as $\rho$ is the charge density related to the electromagnetic four-current $J_\mu = (c \rho , \mathbf{J})$. Similarly, the spatial part $J^\text{V}_l$ represents the current or the flow of the vortex. Together, the density and the current satisfy a continuity equation,
\begin{equation}
 \partial_\lambda J^\text{V}_\lambda = \partial_t J^\text{V}_t + \partial_l J^\text{V}_l = 0.
\end{equation}
The density of vorticity can only decrease (increase) when the vortex flows away from (towards) the current location. In terms of Eq. \eqref{eq:vortex delta function}, the continuity equation is a conservation law or integrability condition, signalling that vortex world lines cannot begin or end in the middle of a superconductor. It is revealing  to regard  $J^\text{V}_\lambda$ as representing the world line of a vortex \emph{itself} (Fig. \ref{fig:single vortex world line}). At each point in spacetime $x$, $J^\text{V}_\lambda(x)$ is the line element of the world line; its temporal component reflects the density (in multiples of $2\pi$) and the  spatial components indicate in which direction the vortex is moving.

With the physical meaning of  the mathematical definition Eq.  \eqref{eq:2+1D vortex current definition} being explicit, we are now ready to generalize to 3+1 dimensions. In three spatial dimensions the topological defects are vortex lines. In spacetime such a line traces out a world sheet. In direct analogy with  $J^\text{V}_\lambda(x)$ being understood as the line element of a vortex world line, we now define,
\begin{equation}\label{eq:4D vortex current}
 J^\text{V}_{\kappa\lambda} (x) = \epsilon_{\kappa \lambda \nu \mu} \partial_\nu \partial_\mu \varphi(x) = 2\pi N \delta^{(2)}_{\kappa\lambda}(x),
\end{equation}
as the \emph{surface element} of the vortex world sheet (Fig. \ref{fig:single vortex world sheet}). In four space-time dimensions, the Levi-Civita symbol $\epsilon_{\kappa \lambda \nu \mu}$ has four indices. The vortex current is antisymmetric in its indices $\kappa\lambda$, which should be the case since those indices can never be parallel when we are to define a surface element. The delta-function is in the plane perpendicular to the world sheet.

What  is the physical meaning of the six independent components of $J^\text{V}_{\kappa\lambda}$? Analogous to the situation in 2+1 dimensions, the temporal components $J^\text{V}_{tl}$ represents the density of vorticity of a vortex line along the $l$-direction, while $J^\text{V}_{kl}$ is the flow of vorticity in the $k$-direction of a vortex line along the $l$-direction. Since vorticity is topologically conserved, there cannot be any vortex flow in the direction in which it is pointing, hence the antisymmetry in $k$ and $l$. For each spatial direction $l$ there is a continuity equation relating the decrease (increase) in vorticity to its flow away (towards) the current location. This may be summarized as $\partial_\kappa J^\text{V}_{\kappa\lambda} =0$. This generalizes the continuity equation to the flow of strings; notice that  this involves three independent conditions instead of the single condition one finds for particles.
To gain some intuition, imagine a straight vortex line in the $z$-direction, the density of which is given by $ J^\text{V}_{tz}$. It can move in either the $x$- or the $y$-direction, denoted by $J^\text{V}_{xz}$ and $J^\text{V}_{yz}$ respectively.  Together they satisfy the continuity equation,
\begin{equation}
 \partial_t J^\text{V}_{tz} +\partial_x J^\text{V}_{xz} +\partial_y J^\text{V}_{yz} = 0.
\end{equation}

In the following sections it will become clear that treating vortex dynamics in terms of its world sheet dynamics is simple and straightforward, allowing 
for immediate derivations of many standard relations in superconductivity.

\section{The vortex world sheet in relativistic superconductors.}\label{sec:vortex world sheet in superconductors}

We will now show how the vortex world sheet appears from the Ginzburg--Landau equations. In section \ref{sec:electrodynamics of two-form sources}, we shall derive the more generic coupling of a vortex current to electromagnetic fields. 

Since we are pursuing a relativistic treatment, we adopt an action rather than a Hamiltonian approach, where time coordinates and derivatives are treated on equal footing with their spatial counterparts. At any time, the equations familiar from condensed matter can be obtained from the spatial part of the Lagrangian.

Before we write down the partition function let us stress that it may be less familiar to researchers in the field of superconductivity, since it will be fully relativistic. In particular it will have a squared time-derivative, whereas most works start with a single time-derivative term. The latter applies to systems which are diffusion-limited. Of course, in actual superconductors vortices are accompanied by such diffusion processes. However, the relativistic action is necessary to derive the vortex worldsheet. Futhermore processes such as Thomas--Fermi screening are in fact ballistic. Finally the validity of this relativistic approach is verified by the results of Section \ref{sec:vortex electrodynamics}. If one wishes to consider diffusion processes, an appropriate term can be added to the Lagrangian at will.

The partition function associated with the relativistic Ginzburg--Landau action deep in the superconducting state is,
\begin{align}
Z &= \int \mathcal{D} \varphi\, \mathcal{D} A_\mu\, \mathcal{F}(A_\mu)\, \te^{\ti/\hbar\  \int \td^4 x\ \mathcal{L} }, \label{eq:GL partition function}\\
   \mathcal{L} &=  -\frac{1}{4\mu_0} F_{\mu\nu}^2 - \frac{\hbar^2}{2m^*} \rho_\text{s} (\partial^\text{ph}_\mu \varphi - \frac{e^*}{\hbar} A^\text{ph}_\mu)^2.\label{eq:GL action}
\end{align}
Here $F_{\mu\nu} = \partial_\mu A_\nu - \partial_\nu A_\mu$ is the electromagnetic field strength; $\mathcal{F}(A_\mu)$ denotes an appropriate gauge fixing condition; $\varphi$ is the superconducting phase related to the order parameter $\Psi = \sqrt{\rho_\text{s}} \te^{\ti \varphi}$; $\rho_\text{s}$ is the superfluid density; $m^*$ and $e^*$ are the mass and charge of a Cooper pair; and most importantly, one must take great care to differentiate between the two velocities in the problem, namely the velocity of light $c$ pertaining to the photon field $A_\mu$, and the phase velocity in the superconductor $c_\text{ph}$. Therefore we have defined $\partial_\mu = (\partial_0 , \nabla)$, $\partial_0 =\frac{1}{c}\partial_t$ and $\partial^\text{ph}_\mu = (  \partial^\text{ph}_0 , \nabla)$, $\partial^\text{ph}_0 =\frac{1}{c_\text{ph}} \partial_t$. Furthermore $A_\mu = (-\frac{1}{c}V, \mathbf{A})$ and $A^\text{ph}_\mu = (-\frac{1}{c_\text{ph}}V, \mathbf{A})$. The last form is dictated by gauge invariance of the second term in Eq. \eqref{eq:GL action}. The metric is $\eta^{\mu\nu} = \text{diag}(-1,1,1,1)$, but we shall use only lower indices for notational simplicity.

We shall for the moment proceed in the relativistic limit where $c_\text{ph} = c$, for simplicity. The equations of motion then follow from variation with respect to $A_\nu$,
\begin{equation}
 \partial_\mu \frac{\partial \mathcal{L}}{\partial (\partial_\mu A_\nu)} - \frac{\partial \mathcal{L}}{\partial A_\nu} = -\frac{1}{\mu_0}\partial_\mu (\partial_\mu A_\nu - \partial_\nu A_\mu) - \frac{\hbar^2}{m^*}\rho_\text{s} \frac{e^*}{\hbar} (\partial_\nu \varphi - \frac{e^*}{\hbar} A_\nu) = 0.
\end{equation}
Now we act with $\epsilon_{\kappa\lambda\rho\nu} \partial_\rho$ on this equation, which leads to,
\begin{equation}\label{eq:relativistic SC EoM}
 - \lambda^2 (\epsilon_{\kappa\lambda\rho\nu} \partial_\mu^2 \partial_\rho A_\nu - \cancel{\epsilon_{\kappa\lambda\rho\nu } \partial_\rho \partial_\nu} \partial_\mu A_\mu ) + \epsilon_{\kappa\lambda\rho\nu} \partial_\rho A_\nu = \frac{\hbar}{e^*} \epsilon_{\kappa\lambda\rho\nu} \partial_\rho \partial_\nu \varphi =\frac{\hbar}{e^*} J^\text{V}_{\kappa\lambda}.
\end{equation}
Here we have defined the London penetration depth $\lambda = \sqrt{\frac{m^*}{\mu_0 {e^*}^2 \rho_\text{s}}} $; the second term vanishes because the antisymmetric contraction of two derivatives; and on the right-hand side we recognize from Eq. \eqref{eq:4D vortex current} the definition of the  vortex current $J^\text{V}_{\kappa\lambda}$. Let us consider the special case $\kappa = t$, and use the definition of the magnetic field $B_l = \epsilon_{lrn} \partial_r A_n$,
\begin{equation}\label{eq:magnetic Meissner with vortex source}
 - \lambda^2 \partial_\mu^2 B_l + B_l = \frac{\hbar}{e^*} J^\text{V}_{tl} =  \frac{\hbar}{e^*} 2\pi N\delta^{(2)}_l (\mathbf{x} ).
\end{equation}
Here we have used relation Eq.  \eqref{eq:4D vortex current}. This is precisely the textbook equation for the Meissner screening of a vortex source of strength $N$, with flux quantum $\Phi_0 = 2\pi \hbar/e^*$ \cite[eq.(5.10)]{Tinkham96}. But instead of ad hoc inserting the delta-function source, we actually derived it from the singular phase field. The only difference is that here also the dynamics are taken into account via the double time derivative contained in $\partial_\mu^2 $. The true power of the vortex world sheet shows itself when considering the electric field $\mathbf{E} = - \nabla A_0 - \partial_t \mathbf{A}$ and the spatial components $J^\text{V}_{kl}$ of the vortex field. This will be further elaborated on in Section \ref{sec:vortex electrodynamics}. But let us first analyze how two-form sources couple to electromagnetism in general, followed by a more general derivation of the above relations invoking a duality mapping, by which we can treat the vortex fields in the action itself, rather than only in the equations of motion. This can be regarded as revealing the more fundamental structure of the problem. The reader who is less interested in these theoretical matters may skip ahead directly to Section \ref{sec:vortex electrodynamics}.

 \section{Electrodynamics of two-form sources.}\label{sec:electrodynamics of two-form sources}
We will formulate here the  generalization of the standard Maxwell action and equations of motion when the sources are not monopoles with charge density $\rho$ and current $J_m$, collected in a vector field $J_\mu = (c\rho , J_m)$, but  instead (vortex) lines with line densities $J_{tl}$ and line currents $J_{kl}$ (which denote the current in direction $k$ of a line that extends in direction $l$), collected in a two-form field $J_{\kappa\lambda} = (J_{tl} , J_{kl}) $. Let us first recall the established knowledge for ordinary electromagnetism, in terms suited for this generalization. For clarity  reasons we again use a shorthand 
notation where we are  intentionally sloppy with contra- and covariant indices, leaving out dimensionful parameters in order to maximally expose the principles.   In the next section we will present the final results that are accurate in this regard.

Let us start considering  a set of electrical monopole sources collected in a source field $J_\mu$ as in the  above,  satisfying a continuity equation/conservation law $\partial_\mu J_\mu = 0$. These sources interact via the exchange of gauge particles, as gauge fields $A_\mu$ that couple locally to the source fields, by an interaction term in the Lagrangian of the form $A_\mu J_\mu$. Because of current conservation, any transformation of the gauge field $A_\mu \to A_\mu + \partial_\mu \varepsilon$, where $\varepsilon$ is any smooth scalar field, will leave the coupling term invariant. Indeed,
\begin{equation}
 A_\mu J_\mu \to A_\mu J_\mu + (\partial_\mu \varepsilon) J_\mu =  A_\mu J_\mu -  \varepsilon \partial_\mu J_\mu =  A_\mu J_\mu.
\end{equation}
Here we performed a partial integration in the second step. The field strength $F_{\mu\nu} = \partial_\mu A_\nu - \partial_\nu A_\mu$ is also invariant under the same gauge transformation. An immediate consequence of this definition are the Bianchi identities or homogeneous Maxwell equations,
\begin{equation}\label{eq:Maxwell Bianchi identities}
 \epsilon_{\alpha \beta\mu\nu} \partial_\beta F_{\mu\nu} =  \epsilon_{\alpha \beta\mu\nu} \partial_\beta \partial_\mu A_\nu = 0,
\end{equation}
because the derivatives commute. These equations comprise $\nabla \cdot \mathbf{B} = 0$ and $\nabla \times \mathbf{E} = - \partial_t \mathbf{B}$. This suggests a Lagrangian of gauge invariant terms,
\begin{equation}
 \mathcal{L}_\text{Maxwell} = -\frac{1}{4} F_{\mu\nu} F_{\mu\nu} + A_\mu J_\mu,
\end{equation}
accompanied by the Euler--Lagrange equations of motion obtained by variation with respect to $A_\nu$,
\begin{equation}
 \partial_\mu F_{\mu\nu} = -J_\nu.
\end{equation}
These are the inhomogeneous Maxwell equations comprising $\nabla \cdot \mathbf{E} = \rho$ and $\nabla \times \mathbf{B} - \partial_t \mathbf{E} = \mathbf{J}$. In a superconductor, one must also add a Meissner term, which in the unitary gauge fix turns into a mass term for the gauge field $A_\mu$,
\begin{equation}\label{eq:Maxwell-Meissner action}
 \mathcal{L}_\text{Maxwell + Meissner} = -\frac{1}{4} F_{\mu\nu} F_{\mu\nu} - \frac{1}{2} A_\mu A_\mu +  A_\mu J_\mu,
\end{equation}
In this form, the Meissner term breaks the gauge invariance of the Lagrangian. This corresponds to releasing the longitudinal degrees of freedom of the photon field. A gauge equivalent perspective is that this degree of freedom represents the phase mode of the superconducting condensate. The equation of motion is modified to,
\begin{equation}\label{eq:EoM Meissner state}
  \partial_\mu F_{\mu\nu} -  A_\nu  = -J_\nu.
\end{equation}

Let us now repeat this procedure for antisymmetric two-form sources $J_{\kappa\lambda} = (J_{tl} , J_{kl}) $. These must obey the continuity equations/conservation laws $\partial_\kappa J_{\kappa\lambda} =0$, reflecting that the density of the source can only increase (decrease) when it flows into (out of) the region under consideration, and that vortex lines cannot end within in the system (no monopoles). Consider now that these sources interact by exchanging two-form gauge fields, that we will tentatively denote by $G_{\kappa\lambda}$. Then these gauge fields couple locally to the sources as $G_{\kappa\lambda} J_{\kappa\lambda}$.  These fields   have to transform under gauge transformations as,
\begin{equation}\label{eq:two-form gauge transformations}
 G_{\kappa\lambda} \to G_{\kappa\lambda} + \frac{1}{2}( \partial_\kappa \varepsilon_\lambda -  \partial_\lambda \varepsilon_\kappa),
\end{equation}
where $ \varepsilon_\lambda$ is any smooth vector field, in order to leave the coupling term invariant as required by the current conservation. Indeed,
\begin{equation}
 G_{\kappa\lambda} J_{\kappa\lambda} \to G_{\kappa\lambda} J_{\kappa\lambda} + (\partial_\kappa \varepsilon_\lambda) J_{\kappa\lambda} = G_{\kappa\lambda} J_{\kappa\lambda} - \varepsilon_\lambda  \partial_\kappa J_{\kappa\lambda} = G_{\kappa\lambda} J_{\kappa\lambda} .
\end{equation}
Here we have used the antisymmetry of $J_{\kappa\lambda}$ in the first step, and partial integration in the second. The field strength $H_{\mu \kappa\lambda} =  \partial_{[\mu} G_{\kappa\lambda]} = \partial_\mu G_{\kappa\lambda} + \partial_\lambda G_{\mu\kappa} + \partial_\kappa G_{\lambda\mu}$ is also invariant under these gauge transformations. An immediate consequence of this definition is the Bianchi identity,
\begin{equation}
 \epsilon_{\nu\mu\kappa\lambda} \partial_\nu H_{\mu \kappa\lambda}  = \partial_{[\nu} \partial_\mu G_{\kappa\lambda]} = 0,
\end{equation}
because the derivatives commute. With these definitions, we can write down a gauge invariant Lagrangian,
\begin{equation}\label{eq:two-form action}
\mathcal{L} = - \frac{1}{12} H^2_{\mu \kappa\lambda} + G_{\kappa\lambda}  J_{\kappa\lambda}.
\end{equation}
Note that this Lagrangian is in terms of the dynamic variables $G_{\kappa\lambda}$, which we will see later is the dual of the electromagnetic field strength $F_{\mu\nu}$. In other words, this Lagrangian is in terms of the electric and magnetic fields themselves, rather than the gauge potential $A_\mu$. The equations of motion follow after variation with respect to $G_{\kappa\lambda}$,
\begin{equation}
 \partial_\mu H_{\mu \kappa\lambda} = - J_{\kappa\lambda}.
\end{equation}
Now, in a gauge-invariance breaking medium such as a superconductor, one must add a Higgs or Meissner term to the Lagrangian as,
\begin{equation}\label{eq:two-form action with Meissner}
 \mathcal{L} = - \frac{1}{12} H^2_{\mu \kappa\lambda} - \frac{1}{4} G_{\kappa\lambda}^2 +  G_{\kappa\lambda}  J_{\kappa\lambda}.
\end{equation}
Up to now we have just reviewed the standard derivation of non-compact $U(1)$ two-form gauge theory. Let us now specialize to the case of a vortex line in a superconductor. For such an Abrikosov vortex, we know that the density $J^\text{V}_{tl}$ is proportional to the magnetic field, and that the magnetic field  is parallel to the spatial orientation of the vortex line. In fact, when the magnetic field intensity coincides with the lower critical field $H_{c1}$, the dimensionful vortex density may be denoted as before, combining Eqs. \eqref{eq:4D vortex current} and \eqref{eq:magnetic Meissner with vortex source},
\begin{equation}
 J^\text{V}_{tl} = \Phi_0 \delta^{(2)}_l(\mathbf{r}),
\end{equation}
where $\Phi_0$ is the flux quantum $\frac{h}{e^*}$. Because of these considerations, the vortex line density should couple to the magnetic field $B_l$. The definition of the Maxwell field strength is,
\begin{align}
 F_{tn}  &= E_n &  F_{mn} &= \epsilon_{mnl} B_l,
\end{align}
If we contract the last definition with $\sum_{mn}\epsilon_{tb mn}$, one finds $B_l = \epsilon_{tlmn} F_{mn} \equiv G_{tl}$. Here we introduce the Hodge dual of the Maxwell field strength $G_{\alpha\beta} \equiv \frac{1}{2}\epsilon_{\alpha\beta \mu\nu} F_{\mu\nu}$. Then the coupling of the vortex line density $J^\text{V}_{tl}$ to the magnetic field $B_l$ is written as $G_{tl} J^\text{V}_{tl}$ and generalizes to $G_{\kappa\lambda} J^\text{V}_{\kappa\lambda}$. Therefore, the general two-form gauge field in Eq. \eqref{eq:two-form action} is now identified as the dual Maxwell field strength $G_{\kappa\lambda}$.

This has one immediate astonishing consequence: the Maxwell field strength $F_{\mu\nu}$ itself {\em has now become a gauge field} ! The gauge transformations Eq. \eqref{eq:two-form gauge transformations} correspond to,
\begin{equation}\label{eq:field strength gauge transformations}
 F_{\mu\nu} \to F_{\mu\nu} + \epsilon_{\mu\nu\kappa\lambda} \partial_\kappa \varepsilon_\lambda.
\end{equation}
How does it come about that  these all too physical $F_{\mu\nu}$s have suddenly turned into gauge variant quantities?  The reason is simple although perhaps defeating the physical intuition: in normal matter we always have electric monopole sources $J_\nu$ with the associated equations of motion $\partial_\mu F_{\mu\nu} = -J_\nu$. In the absence of any such sources, these equations reduce to $\partial_\mu F_{\mu\nu} = 0$. Together with the inhomogeneous Maxwell equations $\epsilon_{\alpha \beta\mu\nu} \partial_\beta F_{\mu\nu} =0$, these imply that the field strength cannot be measured at all. It amounts to the Schwinger wisdom that fields which cannot be sourced do not have physical reality. The formal expression of this fact is that \emph{the field strength becomes pure gauge in the absence of monopole sources}.

Another insight is obtained by taking a closer look at the gauge transformations Eq. \eqref{eq:field strength gauge transformations}. For the Bianchi identities in  Eq. \eqref{eq:Maxwell Bianchi identities} these imply,
\begin{align}
 \epsilon_{\alpha \beta\mu\nu} \partial_\beta F_{\mu\nu} &\to \epsilon_{\alpha \beta\mu\nu} \partial_\beta F_{\mu\nu} + \epsilon_{\alpha \beta\mu\nu} \partial_\beta  \epsilon_{\mu\nu\kappa\lambda} \partial_\kappa \varepsilon_\lambda \nonumber\\
 &= \epsilon_{\alpha \beta\mu\nu} \partial_\beta F_{\mu\nu} + (\partial_\alpha \partial_\lambda - \partial^2 \delta_{\alpha\lambda} ) \varepsilon_\lambda.
\end{align}
In other words, the Bianchi identities are not invariant under these gauge transformations! This makes sense: these identities are a direct result of expressing the field strength in terms of a gauge potential $A_\nu$, which of itself has three degrees of freedom (four minus one gauge freedom). The Bianchi identities serve to restrict the six degrees of freedom contained in $F_{\mu\nu}$ to the proper number of three. In the derivation of the two-form action Eq.  \eqref{eq:two-form action}, we have not assumed anything about the origin of the two-form field. Next to three physical degrees of freedom, there are three gauge degrees of freedom. Therefore the constraints $ \epsilon_{\alpha \beta\mu\nu} \partial_\beta F_{\mu\nu} =0$ are not strictly enforced, but can always be obtained by a suitable gauge transformation.

We never observe the gauge character of the fields $F_{\mu\nu}$ themselves because the only two-form sources to which this action applies that we know of are Abrikosov vortices in a superconductor. The superconducting matter causes a finite penetration depth $\lambda$ for the fields, which is reflected by the  addition of a Meissner term to the Lagrangian. The gauge-invariant form of this term is known to be,
\begin{equation}\label{eq:two-form Higgs term}
 H_{\kappa\lambda\mu} \frac{1}{\partial^2} H_{\kappa \lambda\mu} = - G_{\kappa\lambda} \frac{ \delta_{\kappa\mu} \partial^2 - \partial_\kappa\partial_\mu}{\partial^2} G_{\kappa\lambda},
\end{equation}
in the same way as one can formally write the Meissner term in Eq. \eqref{eq:Maxwell-Meissner action} as $F_{\mu\nu} \frac{1}{\partial^2} F_{\mu\nu}$. However, since the longitudinal components of $G_{\kappa\lambda}$ are not sourced by the conserved Abrikosov vortices, we are naturally led to the Lorenz gauge condition $\partial_\kappa G_{\kappa\lambda} =0$, and the gauge freedom has been removed.  With this gauge condition Eq. \eqref{eq:two-form Higgs term} reduces to $G_{\kappa\lambda}G_{\kappa\lambda}$, that appears in Eq. \eqref{eq:two-form action with Meissner}. In other words, the superconducting medium forces us to the fixed frame action Eq. \eqref{eq:two-form action with Meissner}.

We end up with the action Eq. \eqref{eq:two-form action with Meissner}, and we now put in dimensionful parameters. Please note that this action is equivalent to the regular action as \eqref{eq:GL action}, but with the important difference that here we work with the dual field strength $G_{\kappa\lambda}$ as the dynamic variable instead of the gauge potential $A_\mu$ .The equations of motion (``Maxwell equations for relativistic vortices'') are now obtained straightforwardly by varying with respect to $G_{\kappa\lambda}$ as,
\begin{equation}\label{eq:dual field strength EoM}
\lambda^2 (\partial^2 G_{\mu\nu} - \partial_\mu \partial_\kappa G_{\kappa\nu} +   \partial_\nu \partial_\kappa G_{\kappa\mu}) -  G_{\mu\nu} = -\frac{\hbar}{e^*} J^\text{V}_{\mu\nu}.
\end{equation}
This is to be compared with Eq. \eqref{eq:EoM Meissner state} and Eq. \eqref{eq:magnetic Meissner with vortex source}. The second and third term can be set to zero by a gauge transformation Eq. \eqref{eq:field strength gauge transformations} or alternatively by invoking the Bianchi identities Eq. \eqref{eq:Maxwell Bianchi identities}. The meaning of these equations is that the two-form source $ J^\text{V}_{\kappa\lambda}$, causes an electromagnetic field $G_{\mu\nu} = \frac{1}{2}\epsilon_{\mu\nu\kappa\lambda}F_{\kappa\lambda}$ that is now Meissner screened over a length scale $\lambda$. The case $\mu = t$, together with the definition $B_n = \frac{1}{2}\epsilon_{nab} F_{ab}$ reduces to  Eq. \eqref{eq:magnetic Meissner with vortex source}. 

Summarizing, we have shown here that  the action Eq. \eqref{eq:two-form action with Meissner} can be postulated, from which the correct equations of motion as introduced in Section \ref{sec:vortex world sheet in superconductors} directly follow, without ever mentioning the gauge potential $A_\mu$. One trades in the Bianchi identities Eq. \eqref{eq:Maxwell Bianchi identities} for a set of gauge transformations Eq. \eqref{eq:field strength gauge transformations}. This action is only meaningful in the absence of monopole sources, but is very appropriate when considering two-form sources such as Abrikosov vortices. In the case that the penetration depth $\lambda$ becomes infinitely large, the field strength $F_{\mu\nu}$ recovers its status as a gauge field. This would correspond to the Coulomb phase of two-form sources, as opposed to the Higgs phase that is always realized in superconductors.

As a final note it should be stressed, that although the vortex source is intrinsically dipolar in nature, the equations stated above are not generally valid for any dipole source. Here, the direction of the vortex line is always parallel to the dipole moment. If one should instead consider for instance a string of ferromagnetic material with moments not along the string, one must revert to the omnipotent regular Maxwell equations. 

For the reader familiar with differential forms, we have included an appendix repeating these considerations in metric-independent language, valid in any spatial dimension higher than 2.

\section{Vortex duality in charged superfluids.}\label{sec:vortex duality in charged superfluids}

We shall now rigorously derive the coupling of Abrikosov vortex sources to the electromagnetic fields, starting from the action describing a superconductor in 3+1 dimensions. This follows the same pattern as the first steps of the well known vortex (or Abelian-Higgs) duality in 2+1 dimensions, where by Legendre transformation the vortex--vortex interactions are expressed in an effective electrodynamics sourced by the vortex sources. In a recent publication we demonstrated how to extend this to 3+1 dimensions \cite{BeekmanSadriZaanen11}, in a language that is very closely related to the present context, also resting on two-form gauge theory. However, in this earlier work we concentrated on the neutral superfluid and here we will go one step further by coupling in electromagnetism and integrating out the condensate gauge fields, ending up with an effective action describing the electromagnetism of vortices.  

Our starting point is the partition function Eq. \eqref{eq:GL partition function}. To keep the equations readable, we will transform to dimensionless units denote by a prime (which we suppress when matters are unambiguous), 
\begin{equation}
 S' = \frac{1}{\hbar} S, \;\; x'_m = \frac{1}{a} x_m, \;\;  t' = \frac{c}{a} t, \;\; A'_\mu = \frac{a e^*}{\hbar} A_\mu, \;\; \rho' = \frac{\hbar a^2}{m^* c} \rho_\text{s} , \;\; \frac{1}{\mu'}= \frac{\hbar}{\mu_0 c {e^*}^2} .
\end{equation}
Here $a$ is a length scale relevant in the system, for instance the lattice constant. We will assume the relativistic limit $c_\text{ph} = c$; later we shall return to dimensionful quantities and it will become clear that the phase velocity is playing an essential role for the description of the non-relativistic vortices. The partition function in these dimensionless units reads,
\begin{align}
Z &= \int \mathcal{D} \varphi\, \mathcal{D} A_\mu\, \mathcal{F}(A_\mu)\, \te^{\ti   \int \td^4 x\ \mathcal{L} }, \label{eq:dimensionless GL partition function}\\
   \mathcal{L} &=  -\frac{1}{4\mu} F^2_{\mu\nu}- \frac{1}{2} \rho (\partial_\mu \varphi - A_\mu)^2.\label{eq:dimensionless GL action}
\end{align}

Now we perform the dualization procedure. A Hubbard--Stratonovich transformation of Eq. \eqref{eq:dimensionless GL partition function} leads to,
\begin{align}
Z &= \int \mathcal{D} w_\mu \, \mathcal{D} \varphi\, \mathcal{D} A_\mu\, \mathcal{F}(A_\mu)\, \te^{\ti  \int \mathcal{L}_\text{dual} }, \label{eq:dual GL partition function}\\
   \mathcal{L}_\text{dual} &=  -\frac{1}{4\mu} F^2_{\mu\nu} + \frac{1}{2\rho} w^\mu w_\mu - w^\mu (\partial_\mu \varphi - A_\mu).
\end{align}
Here $w_\mu$ is the auxiliary variable in the transformation, but it is actually the canonical momentum related to the velocity $\partial_\mu \varphi$, which can be found as,
\begin{equation}
 w_\mu = -\frac{\partial \mathcal{L}}{\partial( \partial^\mu \varphi)} = \rho(\partial_\mu \varphi - A_\mu),
\end{equation}
and is related to the supercurrent as $w_\mu = \frac{e^*}{\hbar} J_\mu$. If one integrates out the field $w_\mu$ from Eq. \eqref{eq:dual GL partition function}, one retrieves Eq. \eqref{eq:dimensionless GL partition function}. In the presence of Abrikosov vortices, the superconductor phase $\varphi$ is no longer everywhere single-valued. Therefore it is separated into smooth and multi-valued parts $\varphi = \varphi_\text{smooth} + \varphi_\text{MV}$. The smooth part can be partially integrated yielding,
\begin{align}
Z &= \int \mathcal{D} w_\mu \, \mathcal{D} \varphi_\text{smooth}\, \mathcal{D} \varphi_\text{MV}\, \mathcal{D} A_\mu\, \mathcal{F}(A_\mu)\, \te^{\ti \int \mathcal{L}_\text{dual} }, \\
   \mathcal{L}_\text{dual} &=  -\frac{1}{4\mu} F^2_{\mu\nu} +  \frac{1}{\rho} w^\mu w_\mu +\varphi_\text{smooth} \partial_\mu w^\mu   - w^\mu \partial_\mu \varphi_\text{MV} +w^\mu A_\mu.
\end{align}
The smooth part can now be integrated out as a Lagrange multiplier turning into the constraint $\partial_\mu w^\mu = 0$, the supercurrent continuity 
equation. This constraint can be explicitly enforced by expressing $w^\mu$ as the curl of a gauge field. In 3+1 dimensions, this gauge field is a two-form field,
\begin{equation}
 w^\mu = \epsilon^{\mu\nu\kappa\lambda} \partial_\nu b_{\kappa\lambda}.
\end{equation}
We can now substitute this expression in the partition function; the integral over the fields $w_\mu$  is replaced by one over $b_{\kappa\lambda}$, as long as we apply a gauge fixing term $\mathcal{F}(b_{\kappa\lambda})$ to take care of the redundant degrees of freedom. Since the gauge field is smooth it can be partially integrated to give,
\begin{align}
Z &= \int \mathcal{D} \varphi_\text{MV}\, \mathcal{D} A_\mu\, \mathcal{F}(A_\mu)\, \mathcal{D} b_{\kappa\lambda}\, \mathcal{F}(b_{\kappa\lambda})\, \te^{\ti \int \mathcal{L}_\text{dual} }, \\
   \mathcal{L}_\text{dual} &=  -\frac{1}{4\mu} F^2_{\mu\nu} +  \frac{1}{\rho} (\epsilon^{\mu\nu\kappa\lambda} \partial_\nu b_{\kappa\lambda})^2   -  b_{\kappa\lambda}\epsilon^{\kappa\lambda\nu\mu} \partial_\nu \partial_\mu \varphi_\text{MV} + b_{\kappa\lambda}  \epsilon^{\kappa\lambda\nu\mu } \partial_\nu A_\mu.
\end{align}
Here we recognize the definition Eq. \eqref{eq:4D vortex current} of the vortex source,
\begin{equation}\label{eq:dimensionless vortex definition}
J^\text{V}_{\kappa\lambda}= \epsilon_{\kappa\lambda\nu\mu} \partial^\nu \partial^\mu \varphi_\text{MV},
\end{equation}
and we have derived  the dual partition function,
\begin{align}
Z &= \int \mathcal{D} J_{\kappa\lambda} \, \mathcal{D} A_\mu\, \mathcal{F}(A_\mu)\, \mathcal{D} b_{\kappa\lambda}\, \mathcal{F}(b_{\kappa\lambda})\, \te^{\ti \int \mathcal{L}_\text{dual} }, \\
   \mathcal{L}_\text{dual} &=  -\frac{1}{4\mu} F^2_{\mu\nu} +  \frac{1}{\rho} (\epsilon^{\mu\nu\kappa\lambda} \partial_\nu b_{\kappa\lambda})^2   -  b^{\kappa\lambda}J^\text{V}_{\kappa\lambda} + b_{\kappa\lambda}  \epsilon^{\kappa\lambda\nu\mu } \partial_\nu A_\mu.
\end{align}
The interpretation is as follows. The vortex sources $J^\text{V}_{\kappa\lambda}$  interact through the exchange of dual gauge particles $b_{\kappa\lambda}$ coding for the long range vortex-vortex interactions mediated by the condensate. The gauge field $b_{\kappa\lambda}$ couples as well to the electromagnetic field $A_\mu$. Integrating out the electromagnetic field will lead to a Meissner/Higgs term $\sim b^2_{\kappa\lambda}$, showing that the interaction between vortices is actually short-ranged in the superconductor. However,  we are instead interested in how the electromagnetic field couples to the vortices themselves. Therefore, we shall integrate out the dual gauge field $b_{\kappa\lambda}$. 

The first step is to complete the square in $b_{\kappa\lambda}$. The kinetic term for $b_{\kappa\lambda}$ is proportional to,
\begin{equation}
 -b_{\kappa\lambda}\epsilon^{\kappa\lambda\mu\nu} \partial_\nu \epsilon_{\rho\sigma \alpha \mu} \partial^\alpha b^{\rho\sigma} = -b_{\kappa\lambda} ( \delta^{\kappa\mu} \partial^2 - \partial^\kappa \partial^\mu) b_{\mu\lambda} \equiv -b_{\kappa\lambda} {\mathcal{G}_0^{-1}}^{\kappa\mu}b_{\mu\lambda}.
\end{equation}
Here ${\mathcal{G}_0^{-1}}^{\kappa\mu}$ is the inverse propagator. However, this expression cannot be inverted (the same problem arises in the quantization of the photon field). We can solve this by imposing the Lorenz gauge condition $\partial^\kappa b_{\kappa\lambda} = 0$. Then the inverse propagator is simply ${\mathcal{G}_0^{-1}}^{\kappa\mu} =  \delta^{\kappa\mu} \partial^2$, and its inverse is ${\mathcal{G}_0}_{\kappa\mu} =  \delta_{\kappa\mu} \frac{1}{\partial^2}$. Now we can complete the square,
\begin{align}
 \mathcal{L}_\text{dual} &= \frac{1}{2} \big(b_{\kappa\lambda} -  \frac{\rho}{\partial^2} J^\text{V}_{\kappa\lambda} + \epsilon_{\kappa\lambda\nu\mu } \partial^\nu A^\mu \big)\big( -\frac{\partial^2}{\rho}\big)\big(b^{\kappa\lambda}  -  \frac{\rho}{\partial^2} {J^\text{V}}^{\kappa\lambda} + \epsilon^{\kappa\lambda\rho \sigma} \partial_\rho A_\sigma \big)\nonumber\\
 &\phantom{= } -\frac{1}{2}\big(- J^\text{V}_{\kappa\lambda} + \epsilon_{\kappa\lambda\nu\mu } \partial^\nu A^\mu\big)\big(- \frac{\rho}{\partial^2}\big) \big(- {J^\text{V}}^{\kappa\lambda} + \epsilon^{\kappa\lambda\nu\mu } \partial_\nu A_\mu\big) -\frac{1}{4\mu} F^2_{\mu\nu} .
\end{align}
Then we shift the field $b_{\kappa\lambda} \to b_{\kappa\lambda} +  \frac{\rho}{\partial^2} J^\text{V}_{\kappa\lambda} - \epsilon_{\kappa\lambda\nu\mu } \partial^\nu A^\mu  $ and integrate it out in the path integral to leave an unimportant constant factor. Expanding the remaining terms leads to,
\begin{align}
 \mathcal{L}_\text{dual} &= \frac{1}{2}J^\text{V}_{\kappa\lambda}\frac{\rho}{\partial^2} {J^\text{V}}^{\kappa\lambda} + \frac{1}{2}  \epsilon_{\kappa\lambda\nu\mu } \partial^\nu A^\mu\frac{\rho}{\partial^2} \epsilon^{\kappa\lambda\rho\sigma } \partial_\rho A_\sigma - \rho J^\text{V}_{\kappa\lambda}\epsilon^{\kappa\lambda\nu\mu } \frac{\partial_\nu}{\partial^2} A_\mu -\frac{1}{4\mu} F^2_{\mu\nu}  \nonumber\\
 &= \frac{1}{2}J^\text{V}_{\kappa\lambda}\frac{\rho}{\partial^2} {J^\text{V}}^{\kappa\lambda} - \frac{1}{2}  \rho A^\mu A_\mu - \rho J^\text{V}_{\kappa\lambda}\epsilon^{\kappa\lambda\nu\mu }\frac{ \partial_\nu }{\partial^2} A_\mu -\frac{1}{4\mu} F^2_{\mu\nu}.\label{eq:vortex dynamics action}
\end{align}
In going to the second line we have performed partial integration on the second term and invoked the Lorenz gauge condition $\partial^\mu A_\mu = 0$. We can immediately read off the physics encoded in this action: the first term describes the core energy of the vortices and we shall not need it in this work; the second term is the Higgs mass (including Meissner) for the electromagnetic field; the third term is the coupling term between the electromagnetic field and the vortex source. This last term looks rather awkward  given the derivatives in the denominator. This could signal that the coupling is non-local but that is not the case here. The origin of this coupling follows from the notions presented in section \ref{sec:electrodynamics of two-form sources}: it is not the gauge potential $A_\mu$ but rather the field strength $F_{\mu\nu}$ itself that couples to the vortex source. We can confirm this expectation by computing the equations of motion,
\begin{equation}
 \frac{1}{\mu} \partial_\mu F^{\mu\nu} + \rho \epsilon^{\mu\nu\kappa\lambda} \frac{\partial_\mu}{\partial^2} J^\text{V}_{\kappa\lambda} - \rho A^\nu =0.
\end{equation}
Acting with $\epsilon_{\alpha \beta \gamma \nu} \partial^\gamma$ on this equation, one obtains,
\begin{equation}
\frac{1}{\mu \rho} \epsilon_{\alpha \beta \gamma \nu} \partial^\gamma\partial_\mu F^{\mu\nu}+ \epsilon_{\alpha \beta \gamma \nu} \epsilon^{\mu\nu\kappa\lambda}\frac{\partial^\gamma \partial_\mu}{\partial^2} J^\text{V}_{\kappa\lambda} - \epsilon_{\alpha \beta \gamma \nu} \partial^\gamma A^\nu =0
\end{equation}
Using $F_{\mu\nu} = \partial_\mu A_\nu - \partial_\nu A_\mu$ one can see that from the first term only $\epsilon_{\alpha\beta \mu\nu} \partial^2 F^{\mu\nu}$ survives. Also, using $\partial^\kappa J^\text{V}_{\kappa\lambda}=0$ one can see that $\epsilon_{\alpha \beta \gamma \nu} \epsilon^{\mu\nu\kappa\lambda}\partial^\gamma \partial_\mu J^\text{V}_{\kappa\lambda} =\partial^2 J^\text{V}_{\alpha\beta}$, cancelling the derivatives in the denominator. Altogether we find,
\begin{equation}\label{eq:relativistic vortex equation of motion}
 \frac{1}{2\mu\rho}\epsilon_{\alpha \beta \mu \nu} \partial^2 F^{\mu\nu}  - \frac{1}{2}\epsilon_{\alpha \beta \mu \nu} F^{\mu\nu} = - J^\text{V}_{\alpha\beta}.
\end{equation}
This is the same result as Eq. \eqref{eq:dual field strength EoM}. As we announced earlier, we have derived here  with a completely controlled procedure the dimensionless version of Eq. \eqref{eq:magnetic Meissner with vortex source}, describing the interactions between the vortices and electromagnetic fields inside a relativistic superconductor. Departing from this result we will derive in the next section various physical consequences.  Summarizing this section,  by dualizing  the Ginzburg--Landau action for the superconductor, Eq. \eqref{eq:dimensionless GL partition function} got reformulated in terms of the vortex currents Eq.  \eqref{eq:dimensionless vortex definition} as the active degrees of freedom, that interact via the effective gauge fields parametrizing the rigidity of the superconductor. The latter  were integrated out to obtain the direct coupling of the vortices to the electromagnetic field, leading eventually to the concise equations of motion Eq. \eqref{eq:relativistic vortex equation of motion}.  Although this strategy is well known dealing with vortex `particles' in 2+1 dimensions we are not aware that it was ever explored in the context of the electrodynamics of vortices in  3+1D.  Surely, the derivation presented in the above is in regard with its rigour and completeness strongly contrasting with the rather ad hoc way that the problem is addressed in the standard textbooks  \cite[eq.(5.13)]{Tinkham96}.

\section{Vortex electrodynamics.}\label{sec:vortex electrodynamics}

In order to establish contact with the physics in the laboratory all that remains to be done is to break the Lorentz invariance, doing justice to the fact that the phase velocity of the superconductor as introduced in the first paragraphs of Section  \ref{sec:vortex world sheet in superconductors} is of order of the Fermi velocity of the metal and thereby a tiny fraction of the speed of light. Subsequently we will analyze what the physical ramifications are of our ``Maxwell equations for vortices''.  

The non-relativistic version of the vortex action Eq. \eqref{eq:vortex dynamics action} is,
\begin{align}
 \mathcal{L} &= \frac{\hbar^2}{2m^*}\rho_s J^\text{V}_{tl} \frac{1}{-1/c_\text{ph}^2\partial_t^2 + \partial_k^2} J^\text{V}_{tl}
 -\frac{\hbar^2}{2m^*}\rho_s J^\text{V}_{kl} \frac{c_\text{ph}^2}{-1/c_\text{ph}^2\partial_t^2 + \partial_k^2} J^\text{V}_{kl}\nonumber\\
 &-\frac{{e^*}^2}{2m^* c_\text{ph}^2} \rho_s V^2 -\frac{{e^*}^2}{2m^*} \rho_s  A_m^2\nonumber\\
 &-\frac{e^* \hbar}{m^*} \rho_s \frac{1}{-\frac{1}{c_\text{ph}^2} \partial_t^2 + \partial_k^2} \big[ \frac{1}{c_\text{ph}} J^\text{V}_{ab} \epsilon_{abtm} (\partial_t A_m + \partial_m V) + \frac{1}{2}J^\text{V}_{ta} \epsilon_{tamn} \partial_m A_n \big]\nonumber\\
 &+\frac{1}{2\mu_0c^2} (\partial_t A_n + \partial_n V)^2 - \frac{1}{4\mu_0} (\partial_m A_n - \partial_n A_m)^2.\label{eq:non-relativistic vortex action}
 \end{align}
Varying with respect to $A_\nu$ and acting with $\epsilon_{\alpha \beta \gamma \nu} \partial^\gamma$ and imposing current conservation $\partial^\kappa J^\text{V}_{\kappa\lambda}=0$  will lead to the correct equations of motion. However, the easiest way to obtain the non-relativistic versions of the equations of motion Eq. \eqref{eq:relativistic SC EoM} is to vary Eq. \eqref{eq:GL action} directly with respect to $V$ and $A_n$ respectively,
\begin{align}
- \frac{c^2_\text{ph}}{c^2} \lambda^2 \partial_n E_n - V &= \frac{\hbar}{e^*} \partial_t \varphi,\\
-\lambda^2 \frac{1}{c^2} \partial_t E_n + \lambda^2 \epsilon_{nmk}\partial_m B_k + A_n &= \frac{\hbar}{e^*} \partial_n \varphi.
\end{align}
Here $\lambda = \sqrt{\frac{m^*}{\mu_0 {e^*}^2 \rho_\text{s}}}$ is the London penetration depth. Now we operate on the first equation by $\partial_m = \frac{1}{2}\epsilon_{mtab}\epsilon_{abrt} \partial_r$, and on the second by $\delta_{mn} \partial_t = \frac{1}{2}\epsilon_{tmab}\epsilon_{abtn} \partial_t$ and $\epsilon_{tamn} \partial_m$ respectively to obtain,
\begin{align}
 - \frac{c^2_\text{ph}}{c^2} \lambda^2  \partial_m \partial_n E_n - \partial_m V &= \frac{\hbar}{e^*} c_\text{ph} \frac{1}{2} \epsilon_{mtab} J^\text{V}_{ab},\\
  - \lambda^2 \frac{1}{c^2}\partial^2_t E_m - \lambda^2 \partial_n^2 \partial_t A_m + \partial_t A_m &= \frac{\hbar}{e^*} c_\text{ph}\frac{1}{2} \epsilon_{tmab} J^\text{V}_{ab},\\
  \lambda^2(\nabla^2 - \frac{1}{c^2}\partial_t^2 ) B_a - B_a &= - \frac{\hbar}{e^*} J^\text{V}_{ta}.\label{eq:magnetic equation}
\end{align}
  For the last equation we used the Maxwell  equations $\nabla \times \mathbf{E} = -\partial_t \mathbf{B}$ and $\nabla \cdot \mathbf{B} = 0$. It is equal to the one we found before in Eq. \eqref{eq:magnetic Meissner with vortex source}, obviously, since there the temporal terms do not play a role. 
  
For the equations for the electric field is it useful to choose the Coulomb gauge $\nabla \cdot \mathbf{A} = 0$, and separate the electric field in longitudinal and transversal parts: $\mathbf{E} = \mathbf{E}^\text{L} + \mathbf{E}^\text{T}$, where $\nabla \times \mathbf{E}^\text{L} =0$ and $ \nabla \cdot\mathbf{E}^\text{T} =0$. In the Coulomb gauge we see from the definition $\mathbf{E} = - \nabla V - \partial_t \mathbf{A}$ that $\mathbf{E}^\text{L} = - \nabla V$ and $\mathbf{E}^\text{T} = -\partial_t \mathbf{A}$. We can subtract the first equation above from the second to obtain,
\begin{equation}
 \lambda^2\big( - \frac{1}{c^2} \partial_t^2 E_m +  \nabla^2E^\text{T}_m + \frac{c^2_\text{ph}}{c^2}\nabla^2 E^\text{L}_m \big) - E_m = \frac{\hbar}{e^*} c_\text{ph} \epsilon_{tmab} J^\text{V}_{ab}.\label{eq:electric equation}
\end{equation}

Hence, as in the case of the Maxwell theory for non-relativistic matter one finds instead of the highly symmetric relativistic result Eq. \eqref{eq:relativistic SC EoM}  two equations of motion that are representing the spatial- (magnetic)  and temporal (electrical) sides of the physics, Eq. \eqref{eq:magnetic equation} and Eq. \eqref{eq:electric equation}. One notices that the first `magnetic' equation is quite like the relativistic one while the `electrical' equation is now more complicated for reasons that will become clear in a moment. 

The factor $c_\text{ph}$ on the right-hand side of the electric equation is due to our convention of rescaling the time derivative to having units of $1/\text{length}$ in the definition of $J^\text{V}_{\kappa\lambda}$. Thus all components of $J^\text{V}_{\kappa\lambda}$ have dimensions of a surface density, and multiplying by a velocity is necessary to end up with a current density.  The sign difference on the right-hand side between the electric and magnetic equations is related to the continuity equation $\frac{1}{c_\text{ph}} \partial_t J^\text{V}_{tn} = - \partial_m J^\text{V}_{mb}$. 

To grasp the content of these equations, one should compare the magnetic equation Eq.  \eqref{eq:magnetic equation} with the standard form\cite[eq.(5.13)]{Tinkham96},
\begin{equation}
 \lambda^2\nabla^2 B_a -  B_a = - \Phi_0 \delta^{(2)}_a(\mathbf{r}),\label{eq:magnetic equation Tinkham} 
\end{equation}
Here $\Phi_0 = 2\pi \hbar/e^*$ is the flux quantum. The factor of $2\pi$ is associated with the definition of $J^\text{V}$ as in Eq. \eqref{eq:4D vortex current}. Our treatment automatically takes dynamics into account in the form of temporal derivatives. Otherwise, the correspondence is complete. We have indeed exactly recovered the well-established vortex equation of motion.

The equation for the electric field \eqref{eq:electric equation} looks more involved, but this can be made more insightful by writing the equations for the longitudinal and transversal parts separately,
\begin{align}
 \lambda^2\big( (\frac{c^2_\text{ph}}{c^2}\nabla^2 - \tfrac{1}{c^2}\partial^2_t) E^\text{L}_m - E^\text{L}_m = \frac{\hbar}{e^*} c_\text{ph} \epsilon^\text{L}_{tmab} J^\text{V}_{ab},\label{eq:longitudinal electric equation}\\
\lambda^2\big( (\nabla^2 - \tfrac{1}{c^2}\partial^2_t) E^\text{T}_m - E^\text{T}_m = \frac{\hbar}{e^*} c_\text{ph} \epsilon^\text{T}_{tmab} J^\text{V}_{ab}.\label{eq:transversal electric equation}
\end{align}
The labels on the $\epsilon$-symbol denote that they include a longitudinal or transversal projection. 

We can now read off the following physical relations:

{\bf 1.} \emph{Meissner screening}: from Eq.  \eqref{eq:magnetic equation} in the static limit $\partial_t \to 0$, a vortex line sources a magnetic field, that falls off in the superconductor with a length scale $\lambda$, the familiar Meissner effect.

{\bf 2.} \emph{Thomas--Fermi screening}: from Eq. \eqref{eq:longitudinal electric equation} one infers that the longitudinal (electrostatic) electric field penetrates up to a much smaller length $\frac{c_\text{ph}}{c} \lambda$, which is the Thomas--Fermi length ($c \approx 300c_\text{ph}$). This just amounts to the well known fact that the electrical screening is the same in the metal as in the superconductor.   Notice that this length scale is obtained without referral to the electrons in the normal metal state as in the textbook derivation. 

{\bf 3.} \emph{Dynamic Meissner screening or the Higgs mass}: taking into account the time-dependence, Eq. \eqref{eq:magnetic equation} and Eq. \eqref{eq:transversal electric equation} show that the transversal photon parts of the fields are screened not only in space, but also in time with characteric time scale $\frac{\lambda}{c}$. This is just the familiar statement that the two propagating photon polarizations in 3+1 dimensions acquire a ``Higgs mass''  $\sim \frac{ \hbar}{\lambda c}$ inside the superconductor.

\begin{figure}
 \hfill
 \psfrag{B}{{$\mathbf{B}$}}
 \subfigure[{\scriptsize A vortex in a Josephson junction between two superconductors (grey); it has no normal core. The magnetic field $\mathbf{B}$ is along the vortex; any electric field across the junction causes the vortex to move in the perpendicular direction. Such motion induces electromagnetic radiation that may escape to the outside world.}]{\includegraphics[height=4cm, trim =-2cm 0cm -2cm 0cm ,clip]{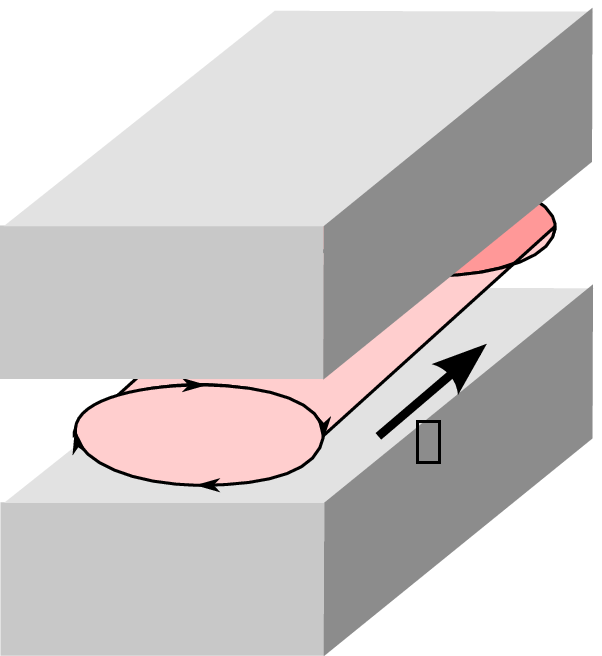}\label{fig:Josephson vortex}}
 \hfill
 \psfrag{B}{{$\mathbf{B}$}}
 \psfrag{E}{{$\mathbf{E}$}}
 \psfrag{v}{{$\mathbf{v}$}}
  \subfigure[Vortex. {\scriptsize Geometry of the electric field $\mathbf{E}$ generated by a vortex line parallel to the magnetic field $\mathbf{B}$ and moving with a speed $\mathbf{v}$. This phenomenon related to the Lorentz force follows directly from the vortex equations of motion.}]{\includegraphics[height=4cm]{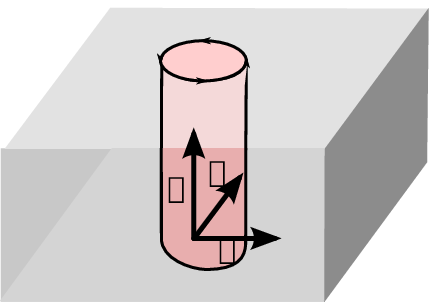}\label{fig:vortex motion}}
 \hfill
\caption{Additional vortex configurations. }
\end{figure}

{\bf 4.} \emph{Electrical field of a moving vortex and the Nernst effect}: disregarding the dynamical term in Eq. \eqref{eq:electric equation}, one is left with
\begin{equation}
 E_m = -\frac{\hbar}{e^*} c_\text{ph} \epsilon_{tmkl} J^\text{V}_{kl}.
\end{equation}
Recall from section \ref{sec:vortex fields} that we had interpreted $ J^\text{V}_{kl}$ as the flow or velocity in the $k$-direction of a vortex line in the $l$ direction. Since we know that one vortex line carries a magnetic flux of $\Phi_0 = 2\pi \frac{\hbar}{e^*}$, we can write $\frac{\hbar}{e^*} c_\text{ph} J^\text{V}_{kl} = v_l B^0_k$, where $B^0$ denotes the field associated with one quantum of flux, and $v_l = c_\text{ph} \hat{e}_l$ is the velocity. In practice there is always a drag force that greatly slows down the vortices. Still, Josephson vortices that do not have a normal core (Fig. \ref{fig:Josephson vortex}) may achieve this large speed. With this interpretation, \eqref{eq:electric equation} reads,
\begin{equation}
 \mathbf{E} = - \mathbf{v} \times \mathbf{B}^0, 
\end{equation}
which is precisely the known result \cite{Josephson65} for the electric field generated by a  vortex moving in a magnetic field $B^0$ (Fig. \ref{fig:vortex motion}). When the  motion is  caused by  a temperature gradient this is responsible for the large Nernst effect of the vortex fluid. 

{\bf 5.} \emph{AC Josephson relation}: another interpretation of Eq. \eqref{eq:longitudinal electric equation} is found by inserting the definition of the vortex current, $J^\text{V}_{ab} = \epsilon_{abtn} \frac{1}{c_\text{ph}} \partial_t \partial_n \varphi$, taking $m$ as the longitudinal direction and neglecting the higher derivative terms. In this case,
\begin{equation}
 \partial_m V =  \frac{\hbar}{e^*} \partial_t (\partial_m \varphi).
\end{equation}
Here the left-hand side is the potential difference, and the right-hand side is the time derivative of the superconducting phase difference. This is exactly the AC Josephson relation. The full equations Eq.  \eqref{eq:electric equation} reveal also that the induced electric field is screened inside the superconductor. 

{\bf 6.} \emph{Moving vortices as radiation sources}:  in the same spirit, the moving vortex is also inducing dynamic transversal fields according to Eq.  \eqref{eq:transversal electric equation}. In other words: moving vortices radiate \cite{BulaevskiiChudnovsky06}. But since the field is Meissner screened, it is very hard to detect this radiation. All our results also apply to Josephson vortices (line vortex solutions in a Josephson junction between two superconductors parallel to the interface, Fig. \ref{fig:Josephson vortex}), which differ only in the regard that they do not have a normal core. There is much recent interest in radation from (arrays of) Josephson junctions, see e.g. \cite{SavelevYampolskiiRakhmanovNori10}. Since inside the junction the field is not expelled by  Meissner  and metallic screening, the radiation may escape to the outside world. In this literature one finds the following result\cite[eq.(13)]{MintsSnapiro95},
\begin{equation}
 - \tilde{\lambda}^2\nabla^2 \mathbf{A} + \mathbf{A} = \frac{\hbar}{e^*} \nabla \phi. 
 \label{radiation}
\end{equation}
Here $\tilde{\lambda}$ differs from $\lambda$ because of a special geometry. Compare this with a result that follows from Eq.  \eqref{eq:transversal electric equation},
\begin{equation}
 \partial_t\big[ -\lambda^2\big( (\nabla^2 - \tfrac{1}{c^2}\partial^2_t) \mathbf{A}  + \mathbf{A} \big] =\partial_t \big[ \frac{\hbar}{e^*}  \nabla \varphi \big],
\end{equation}
confirming Eq. (\ref{radiation}) but showing in addition how to take care of a possible time dependence of the photon field. 

Summarizing, to the best of our knowledge we have addressed all known electrodynamical properties of vortex matter departing from a single action principle. 

\section{Conclusions.}

We are of the opinion that our action principle for vortex electrodynamics Eq. \eqref{eq:vortex dynamics action} resp. \eqref{eq:non-relativistic vortex action} and the associated ``vortex-Maxwell'' equations Eq. \eqref{eq:relativistic vortex equation of motion}, \eqref{eq:magnetic equation} and \eqref{eq:electric equation} deserve a place in the textbooks on the subject. In contrast with the clever but improvising discussions one usually finds, our formulation has the same `mechanical' quality as for instance the Landau--Lifshitz treatise of electromagnetism. One just departs from the fundamentals, to expose the consequences by unambiguous and straightforward algebraic manipulations that are worshipped by any student of physics. A potential hurdle is that one has to get familiar with the two-form gauge field formalism, but then again this belongs to the kindergarten of differential geometry and string theory. 

Our analysis also reveals the origin of the peculiar nature of this vortex electrodynamics. The realization that it is in fact  governed by a two-form gauge structure amounts to an entertaining excursion in the fundamentals of gauge theory itself, nota bene associated with the superficially rather mundane and technology-oriented vortex physics, at least when viewed from the perspective of fundamental physics. In fact, our pursuit started in the quite different subject of the dual description of Bose-Mott insulators in terms of superconducting vortex matter where quite similar issues arise. In this context one does not get anywhere with physical intuition and one just needs full mathematical control to make any progress. Therefore we decided to inspect these matters first in the more familiar present context.   

On the practical side, as we implicitly emphasized in the last section our approach offers a unified description of the electrodynamics of vortices. Although we got as far as recovering the known physical effects in terms of special limits of our equations, we sense that there is a potential to use them to identify hitherto unknown effects and perhaps to arrive at a more complete description of the electrodynamics vortex matter.  Being well aware of the large body of knowledge of this large field in physics, we leave this as an open question to the real experts.  
  
\paragraph{Acknowledgments.}
We thank Peter Kes, Lev Bulaevskii and Yan Liu for useful discussions. This work was supported by the Netherlands foundation for Fundamental Research of Matter (FOM) and the Nederlandse Organisatie voor Wetenschappelijk Onderzoek (NWO) via a Spinoza grant.

\appendix
\section{Electrodynamics with differential forms.}
For the reader familiar with the mathematical language of differential forms, we present the electrodynamics of vortex sources for any dimension $d = D+1$ higher than 2. For our purposes, a differential form can be thought of as something that can be integrated over; in other words: it is a density function combined with the integrand. For instance, the electric field is a 1-form $\mathsf{E} = E_i \td x_i = E_x \td x + E_y \td y + E_z \td z$. Higher forms are always obtained through the wedge product $\mathsf{a} \wedge \mathsf{b}$, which is the antisymmetrization of the tensor product of $\mathsf{a}$ and $\mathsf{b}$. Another common operation is the Hodge dual $*\mathsf{a}$ of $\mathsf{a}$, which turns an $n$-form into a $(d-n)$-form. For instance in three spatial dimensions $*\mathsf{E} = E_x \td y \wedge \td z + E_y \td z \wedge \td x + E_z \td x \wedge \td y$. For a pedagogical introduction to differential forms in Maxwell electrodynamics see \cite{WarnickRusser06}.

In the familiar case of $d=3+1$, a 1-form is a line density or ``field intensity'' like the electric field; a 2-form is a surface density or flux density like the magnetic field; a 3-form is a volume density like the charge density. Confusion may arise when it is not immediately clear whether an object is an $n$-form or a $d-n$-form, which is important for generalization to other dimensions. We distinguish the regular Hodge dual $*$ from the spatial Hodge dual $*_s$, where the latter does not involve the temporal dimension. The exterior derivative operator is $\mathsf{d} = \frac{\partial}{\partial t}\ \td t \wedge + \sum_i \frac{\partial}{\partial x_i}\ \td x_i \wedge$, and the one with only spatial components is $\mathsf{d}_s = \sum_i \frac{\partial}{\partial x_i}\ \td x_i \wedge$. The Leibniz rule is $\mathsf{d} ( \mathsf{a} \wedge \mathsf{b} ) = \mathsf{d}  \mathsf{a} \wedge \mathsf{b} + (-1)^r  \mathsf{a} \wedge \mathsf{d}\mathsf{b}$, where $\mathsf{a}$ is an $r$-form. This can be used for partial integration.
\begin{table}
 \begin{tabular}{ll@{ }cc@{ }c@{ }p{2.5cm}}
 name& field & ?-form & 2+1d& 3+1d &representative in d=3+1 \\
\hline\\
electric field &$\mathsf{E}$ & 1& 1 & 1 & $E_x\ \td x$ \\
dielectric current & $\mathsf{D} = \varepsilon *_s \mathsf{E}$ & $d$-2 & 1 & 2 & $D_x\ \td y \wedge \td z$\\
magnetic field & $\mathsf{B}$ & 2 & 2 & 2 &$ B_x\ \td y \wedge \td z$ \\
magnetic intensity &$\mathsf{H} = \mu *_s \mathsf{B}$ & $d$-3 & 0 & 1 &$ H_x\ \td x$\\
charge density &$\mathsf{\rho}$ & $d$-1 & 2 & 3 & $\rho\ \td x \wedge \td y \wedge \td z$\\
current density & $\mathsf{J}$ & $d$-2 & 1 & 2 & $J_x\ \td y \wedge \td z$ \\
covariant current & $\mathsf{j} = \mathsf{\rho} + \mathsf{J} \wedge \td t$ & $d$-1 & 2 & 3 & $j_x\ \td y \wedge \td z \wedge \td t$ \\
field strength &$\mathsf{F} = \mathsf{B} + \mathsf{E} \wedge \td t$& 2 & 2 & 2 & $F_{xy}\ \td x \wedge \td y$\\ 
gauge potential & $\mathsf{A}$ & 1 & 1 & 1 & $A_x\ \td x$\\
vortex source & $\mathsf{J}^\text{V}$ & $d$-2 & 1 & 2 & $J^\text{V}_{xy}\ \td x \wedge \td y$\\
Lagrangian density & $\mathsf{\mathcal{L}}$ & $d$ & 3 & 4 & $\mathcal{L}\ \td t \wedge \td x \wedge  \td y \wedge \td z $\\
\\
\hline
 \end{tabular}
\caption{Electrodynamical quantities in differential forms. Here $\varepsilon$ and $\mu$ are the electric permittivity and the magnetic permeability, and $*_s$ is the spatial Hodge dual. Other factors of $c$ are suppressed. Minus signs are subject to convention.}\label{table:electrodynamics differential forms}
\end{table}

In table \ref{table:electrodynamics differential forms} we have listed the differential forms of the relevant fields. Some of these definitions seem perhaps unfamiliar. In particular, we are used to thinking of the magnetic field as a vector field; however, its solenoidal nature is typical of a two-form. This becomes even more clear when it is expressed as the curl of the vector potential $\mathsf{B} = \mathsf{d}_s \mathsf{A}$, which holds in $3+0$ dimensions. Also the current density $\mathsf{J}$ is naturally a flux or a 2-form, but its generalization is as a $d-2$-form. One way to see that this must be so, is to write down the continuity equation in differential forms,
\begin{equation}
 \partial_t \rho + \nabla \cdot \mathbf{J}=0 \qquad \to \qquad (\partial_t \mathsf{\rho} + \mathsf{d}_s \mathsf{J} )\wedge \td t = \mathsf{d}\mathsf{j} = 0.
\end{equation}
The current density appearing as a vector field for instance in Ohm's law, $\mathbf{J} = \sigma \mathbf{E}$ is actually the spatial Hodge dual of $\mathsf{J}$.

We shall now write down the familiar expressions of Maxwell electrodynamics. The Lagrangian density is a spacetime volume density. All terms must therefore combine into $d$-forms. The field strength is $\mathsf{F} = \mathsf{d} \mathsf{A}$. From this definition it is clear that the gauge transformations $\mathsf{A} \to \mathsf{A} + \mathsf{d} \mathsf{\xi}$, with $\mathsf{\xi}$ any 0-form, leave the field strength unchanged, since $\mathsf{d}^2 = 0$. The field strength is contracted with its dual to obtain a $d$-form in the Lagrangian. The sources couple to the gauge potential (this is another reason why the source is a $d-1$ form). The Maxwell action is then,
\begin{equation}
 S = \int -\mathsf{F} \wedge * \mathsf{F} + \mathsf{A} \wedge \mathsf{j}.
\end{equation}
The second term is also invariant under the same gauge transformations, provided that $\mathsf{d} \mathsf{j} = 0$, the continuity equation. The Euler--Lagrange equations are,
\begin{equation}
 \mathsf{d} \frac{\partial \mathsf{L} } {\partial \mathsf{d} \mathsf{A} } - \frac{\partial  \mathsf{L} } {\partial  \mathsf{A} }= 0,
\end{equation}
resulting in the inhomogeneous Maxwell equations,
\begin{equation}
 \mathsf{d} *\mathsf{d} \mathsf{A} =  \mathsf{d} *\mathsf{F} = -\mathsf{j}, \qquad (\partial_\mu F_{\mu\nu} = -J_\nu).
\end{equation}
Applying the exterior derivative on this equation directly leads to the continuity equation, since $\mathsf{d}^2 = 0$. Similarly, from the definition $\mathsf{F} = \mathsf{d} \mathsf{A}$ it immediately follows that,
\begin{equation}
 \mathsf{d} F = 0,
\end{equation}
which are the homogeneous Maxwell equations, or in this context rather the Bianchi identities.

Now let us repeat the reasoning of section \ref{sec:electrodynamics of two-form sources}. In the absence of monopole sources $\mathsf{J}$, we have both $\mathsf{d} *\mathsf{F} = 0$ and $\mathsf{d} \mathsf{F} =0$. This implies that the field strength has become ``pure gauge''. The first of these equations still holds when we add any 1-form $\xi$ as $*\mathsf{F} \to *\mathsf{F}  + \mathsf{d} \mathsf{\xi}$. The original Bianchi identities are not invariant under these transformations. The dual field strength $*\mathsf{F}$ turns into a gauge potential, and is accompanied by its own field strength $\mathsf{K} = \mathsf{d} *\mathsf{F}$, which contracts with its dual in the Lagrangian. The field strength can couple to a $d-2$-form source, which we anticipatingly denote by $\mathsf{J}^\text{V}$, provided that $\mathsf{d} * \mathsf{J}^\text{V} = 0$. Indeed,
\begin{equation}
\mathsf{F} \wedge \mathsf{J}^\text{V} \to \mathsf{F} \wedge \mathsf{J}^\text{V}  + * \mathsf{d} \mathsf{\xi}\wedge \mathsf{J}^\text{V} = \mathsf{F} \wedge \mathsf{J}^\text{V}  - \mathsf{\xi}\wedge   \mathsf{d} * \mathsf{J}^\text{V} =  \mathsf{F} \wedge \mathsf{J}^\text{V}.
\end{equation}
The second step is achieved by partial integration, and the last equality holds if the vortex current is conserved, $\mathsf{d} * \mathsf{J}^\text{V}=0$. The action for vortices directly sourcing the field tensor is, (with $\mathsf{G} = * \mathsf{F}$),
\begin{equation}
 S = \int -\mathsf{K} \wedge * \mathsf{K} + \mathsf{F} \wedge \mathsf{J}^\text{V} = \int -\mathsf{K} \wedge * \mathsf{K} + \mathsf{G} \wedge * \mathsf{J}^\text{V}.
\end{equation}
Variation with respect to $\mathsf{G}$ leads to,
\begin{equation}
 *\mathsf{d} * \mathsf{d} \mathsf{G} =  -\mathsf{J}^\text{V}.
\end{equation}
This equation corresponds to $\epsilon_{\kappa\lambda\mu\nu} \partial^2 F_{\mu\nu}  =  -J^\text{V}_{\kappa\lambda}$ as in eq. \eqref{eq:relativistic vortex equation of motion}, but is valid in any dimension. The addition of a Meissner term results in
\begin{equation}
 S = \int -\mathsf{K} \wedge * \mathsf{K} -  \mathsf{G} \wedge *\mathsf{G} +  \mathsf{G} \wedge * \mathsf{J}^\text{V}.
\end{equation}
and,
\begin{equation}
  *\mathsf{d} * \mathsf{d} \mathsf{G} - \mathsf{G} =  -\mathsf{J}^\text{V}.
\end{equation}
This is the equation of motion for $d-1$-dimensional superconductors, which have $d-2$-dimensional vortex worldbranes $\mathsf{J}^\text{V}$.
\bibliography{references}
\end{document}